\documentclass[runningheads]{llncs}

\usepackage{graphicx}
\usepackage{algorithm2e}
\usepackage{amsmath,amsfonts,amssymb}
\begin{document}
\title{2D Qubit Placement of Quantum Circuits using LONGPATH}
\author{Mrityunjay Ghosh\inst{1,3} \and
Nivedita Dey\inst{2} \and
Debdeep Mitra\inst{2} \and
Amlan Chakrabarti\inst{3} }
\authorrunning{M. Ghosh et al.}
\institute{Department of Computer Science and Engineering, Amity University, Kolkata
\and
Department of Computer Science and Engineering, University of Calcutta \and
A. K. Choudhury School of Information Technology, University of Calcutta\\
}
\maketitle

\begin{abstract}
In order to achieve speedup over conventional classical computing for finding solution of computationally hard problems, quantum computing was introduced. Quantum algorithms can be simulated in a pseudo quantum environment, but implementation involves realization of quantum circuits through physical synthesis of quantum gates. This requires decomposition of complex quantum gates into a cascade of simple one qubit and two qubit gates. The methodological framework for physical synthesis imposes a constraint regarding placement of operands (qubits) and operators. If physical qubits can be placed on a grid, where each node of the grid represents a qubit then quantum gates can only be operated on adjacent qubits, otherwise SWAP gates must be inserted to convert non-Linear Nearest Neighbor architecture to Linear Nearest Neighbor architecture. Insertion of SWAP gates should be made optimal to reduce cumulative cost of physical implementation. A schedule layout generation is required for placement and routing apriori to actual implementation. In this paper, two algorithms are proposed to optimize the number of SWAP gates in any arbitrary quantum circuit. The first algorithm is intended to start with generation of an interaction graph followed by finding the longest path starting from the node with maximum degree. The second algorithm optimizes the number of SWAP gates between any pair of non-neighbouring qubits. Our proposed approach has a significant reduction in number of SWAP gates in 1D and 2D NTC architecture.

\end{abstract}

\section{Introduction}
Quantum Computing is a new computational paradigm to demonstrate the exponential speedup over classical non-polynomial time algorithms. Here probability and uncertainty replace determinism, in which energy can be delivered in discrete packets called quanta exhibiting dual nature to remain in particle form and also in wave form. Intrinsic features of quantum states like superposition and entanglement have made the system and its components fragile because whenever they interact with the environment, the information stored in the system decoheres thus resulting in error and consequent failure of computation \cite{MeierSpinCluster}. To overcome the debilitating effects of decoherence and realize subtle interference phenomena in systems with many degrees of freedom, reliability should be enhanced by encoding a given computational state using blocks of quantum error correcting code. Basic design principle of a fault-tolerant protocol is to avoid spreading out of a single qubit error due to fault gate or noise on a quiescent qubit to remaining qubits within one block of QECC \cite{HollenbergQuasi2D} \cite{MaslovPlacement}. Fault-tolerant quantum error correction techniques include Shor fault-tolerant error correction, Steane error-correction, Knill error-correction etc. 

FT Quantum gates (single qubit or multiple qubit) are restrictive as they can only be applied on physically adjacent qubits. Various quantum techniques have been proposed for enabling various degrees of qubit interactions. 1D architectures are highly restrictive, since it can access only two neighbours per qubit, 2D architectures enable a qubit except for qubits present at the boundaries to access four adjacent neighbours and 3D architectures with six neighbours per qubit which has highly complex access method. Ion trap technology \cite{WhitneyAutomatedGeneration} uses 1-D interaction. Quantum dot (QD), superconducting (SC), neutral atom (NA) and photonics use 2-D interaction\cite{PAQCS}. Cubic lattice crystal architecture in cellular automata uses 3-D interaction sequence of SWAP operations to couple any two non-adjacent distant qubits increases circuit latency and error rate \cite{PAQCS} \cite{PalerTopological}. Amelioration of error threshold requires intricated control on quantum gate construction with higher fidelities followed by robust QECCs.
In conventional VLSI design, the circuit placement starts with a weighted hypergraph where nodes represent standard cells and hyperedges represent connections among these cells. Circuit placement determines centre positions for nodes with a predefined size such that objective function specific constraints can be optimized. Placement is followed by routing to connect placement cells through wires. Cost of computation in conventional VLSI technique relies upon wirelength, rate of power consumption and circuit delay. VLSI algorithms can be used to embed a weighted undirected interaction graph into a grid. But in qubit placement, positions of qubits keep varying at each itteration of SWAP gate insertion. Dynamic placement algorithms are to be devised so as to tackle time-variant nature of qubits to place it into a grid. By introducing dynamicity in placement, communication can be reduced \cite{PAQCS}. After physically placing qubits in specific grid nodes using SWAP gates, exchange of qubits might be required in an ordered fashion to route two distant non-adjacent qubits towards each other in order to apply a quantum gate \cite{DiVincenzo} \cite{PAQCS}. Which, in turn, will have impact on all other qubits as their positions in placement grid will also be disturbed. Placement solution must be made reversible such that all moved qubits may return to their initial location by applying the same sequence of SWAP gates in reverse order\cite{WilleOptimalSWAP}.
Apart from considering mobility of qubits in placement grid in 2-D architecture, compatibility of $n$-qubit quantum gates should also be kept in mind, as they may not be directly implementable in a physical quantum machine. Consequently, gates must be further decomposed using a set of supported primitive quantum operators in the physical machine description (PMD) of the quantum machine. Realization of quantum gates on different physical quantum machines requires different number of primitive quantum operations. Physical realization of quantum computers is a function of unitary Hamiltonian operator to perform time-evolution of a closed quantum systems. Different quantum systems have different Hamiltonian, subsequently different PMDs. PMDs include Quantum Dot (QD) architecture where a qubit is represented by spin states of two electrons, Superconducting (SC) where qubits are represented by charged carriers, ION trap (IT) where quantum system is based on a 2-D lattice of confined ions each representing a movable physical qubit within the lattice, Neutral atom (NA) where trapped neutral atoms, those are isolated from the environment exploit quantum structure, Linear Photonics (LP) where a probabilistic two-photon gate is teleported into a quantum circuit with high probability and Non-linear Photonics (NP) where quantum system is based on weak cross-Kerr non-linearities etc \cite{OptimizedQuantLib}. There exists a specific compatibility relationship between quantum gates and PMDs, viz. Controlled NOT (CNOT), SWAP gates are supported in LP system, whereas SWAP gate is not available in NP physical machine description (PMD). \cite{PAQCS} \cite{OptimizedQuantLib}

\begin{figure}
\begin{center}
\includegraphics[width=200pt]{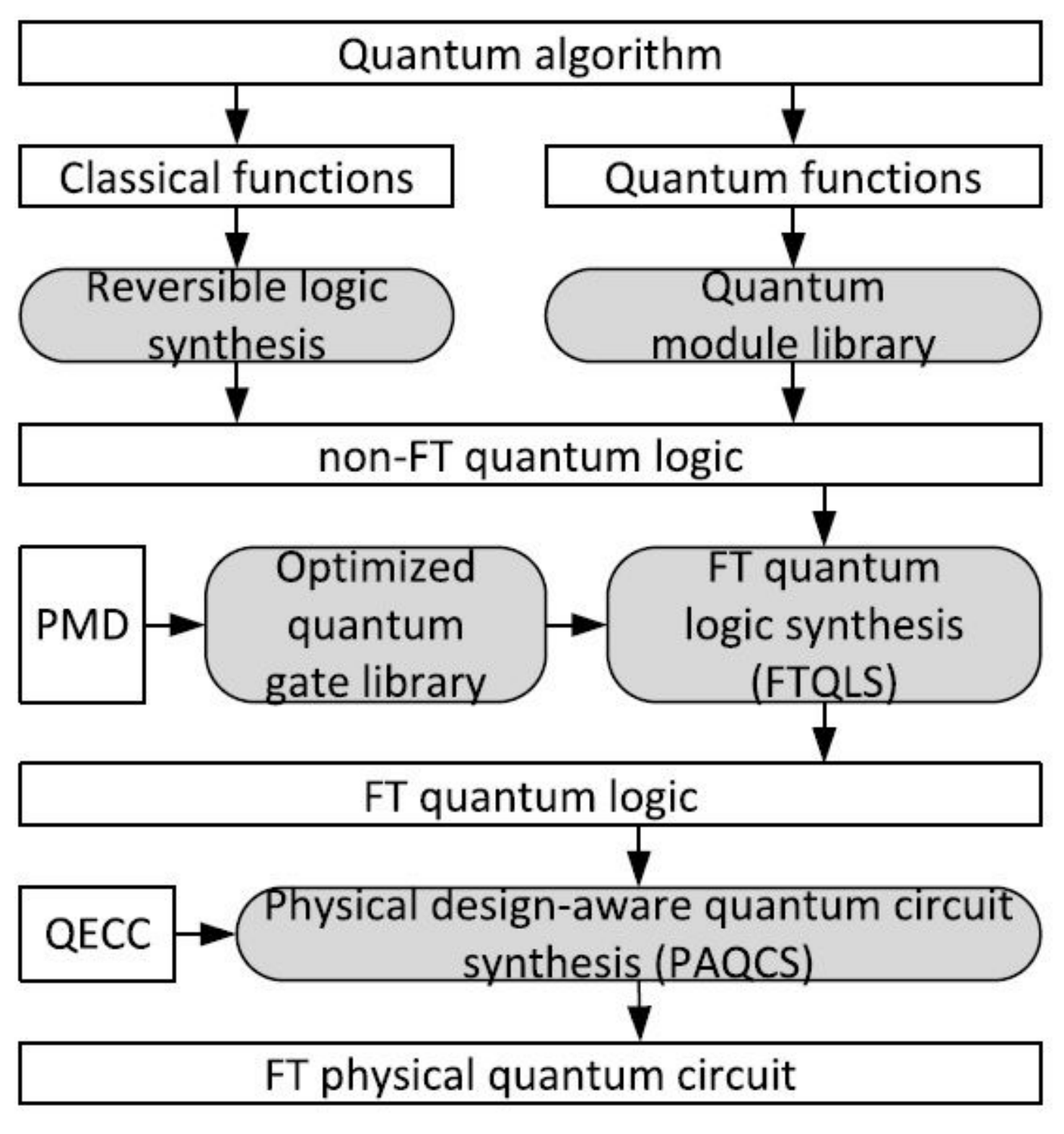}
\end{center}
\caption{Synthesis flow of quantum circuits \cite{PAQCS}} \label{SynthesisFlow}
\end{figure}

The complexity involved in the physical synthesis of quantum circuits can be effectively reduced by decomposing the overall synthesis into these stages \cite{DiVincenzo}. Initial stage starts with a quantum algorithm containing both classical and quantum functions. Arbitrary quantum functions are difficult to synthesized, but they can be synthesized with the help of a quantum module Library QLib. Classical reversible functions can be synthesized using reversible logic synthesizer, whereas QLib is helpful to convert high-level quantum logic gates into low-level primitive quantum gates as it contains many commonly used quantum modules \cite{FalconerMosca}. Quantum Gate library can be optimized in terms of number of primitive-quantum-operations and associated delay by exploring one-qubit and two-qubit identity rules to remove redundancies in quantum gate implementation \cite{PAQCS}.

The circuit thus obtained after first stage is optimized in the next stage using an Fault-Tolerant Quantum Logic Synthesis (FTQLS) which synthesizes and optimizes the non FT logic to FT logic circuits.
The last stage involves a direct synthesis of physical circuit by placing the qubits on a 2-D grid and routing cells properly to reduce communication overhead. In this stage, physical cost of implementing QECC is considered, as optimized cost will be a function of chosen QECC and corresponding PMD \cite{PAQCS}.
The remainder of this paper is as follows. Section II presents the earlier work related to physical design and synthesis of quantum circuits. Section III presents physical design based problem statement as well as different approaches towards the problem. In section IV we have proposed our approach and a novel algorithm for optimizing SWAP gates in quantum circuits. Section V describes complexity analysis of our algorithm and last section will be the concluding part of this paper.

\section{Literature Review}
Logical design phase of quantum circuits, in algorithmic level assumes position independent interaction of qubits. But, physical design phase, in implementation level relies on neighbouring-qubit interactions \cite{FalconerMosca}. Y.Hirata M.Nakanishi and S.Yamashita first proposed an efficient method to convert an arbitrary quantum circuit to one on an LNN architecture applying permutation circuit for each qubit ordering \cite{HirataNakanishi}.

Later, a trade-off between scalability and complexity is proposed by M.Saeedi, R.Wille and R.Drechsler incorporating Nearest Neighbour Cost (NNC) based decomposition methods\cite{SaaediWilleLNN}. A.Shafaei, M.Saeedi, M.Pedram formulated Minimum Linear Arrangement (MINLA) using qubit reordering to improve circuit locality in an interaction graph \cite{ShafaeiSaeediPedramLNN}. Graph partitioning based approach for LNN synthesis was proposed by A.Chakraborty et. al. to provide significant reduction in the number of SWAP gates using reordering of qubit lines in Quantum Boolean Circuits (QBCs). Later, N.Mohammadzadeh et. al. performed quantum physical synthesis applying netlist modifications through scheduled layout generation and iterative update scheduling using a gate-exchanging heuristic \cite{MohammadzadehNetlist}. Later, scientists N.Mohammadzadeh, M.S.Zamani et. al. proposed auxiliary qubit insertion technique after layout generation to meet design constraints using ion trap technology \cite{MohammadzadehPhysicalDesign}. H.Goudarzi et. al. presented a physical mapping tool for quantum circuits using trapped ion as PMD to generate the optimal universal logic block (ULB) which can perform any logical fault-tolerant (FT) quantum operation with minimum latency \cite{GoudarziTrappedIon}. S.Choi and V.Meter first proposed an adder for 2D nearest neighbour with $\theta(\sqrt{n}$) depth and $O(n$) number of qubits \cite{ChoiMeter}. Based on the work presented by Choi and Van Meter, Quantum addition on 2-dimensional nearest-neighbour architecture was proposed by M.Saeedi, A.Shafaei and M.Pedram where modified circuit structures for basic blocks of quantum adder were introduced to provide a significant reduction in communication overhead by adding concurrency to consecutive blocks and also by parallel execution of expensive Toffoli gates. The suggested optimizations, introducing consecutive block architecture, can improve depth from $140\sqrt{n}+K_{1}$ to $92\sqrt{n}+K_{2}$ for constant values of $K_{1}$ and $K_{2}$ \cite{SaeediConstantFactor}. Later, P.Pham and K.M.Svore presented a 2D nearest neighbour quantum architecture for Shor's algorithm to factor an n-bit number in $O(log^{3}n)$ depth. Their proposed circuit incorporating algorithmic improvements (carry-save adders and parallelized phase estimation) and architectural improvements (irregular two-dimensional layouts and constant-depth communication with parallel modules)results in an exponential improvement in nearest-neighbour circuit depth at the cost of a polynomial increase in circuit size and width \cite{PhamSvore}. 

A.Shafaei, M.Saeedi and M.Pedram proposed optimization methods using Mixed Integer Programming (MIP) formulation for time-variant dynamic placement of qubits. This approach places frequently interacting qubits as close as possible on the 2D grid to lessen the requirement of SWAP gates while routing \cite{ShafaeiPlacement2D}. Scalability to a large extent was achieved by Chia-Chun Lin and Susmita Sur-Koley in their work to design an effective physical design-aware quantum circuit synthesis methodology (PAQCS) incorporating quantum error-correcting code where two algorithms are proposed for qubit placement and channel routing respectively. With the help of these two algorithms, the overhead of converting a logical to a physical circuit was reduced by 30.1\%, on an average, relative to previous work \cite{PAQCS}.

\section{Physical Design of Quantum Circuits}
In order to simulate a quantum algorithm physical realization of quantum circuits is required incorporating inherently reversible quantum gates. A reversible function establishes an one-to-one correspondence between input and output assignment where same number of variables are there in domain and range set. A circuit realizing a reversible function is a cascade of reversible quantum gates. Two common reversible gates include
Controlled-Controlled NOT (Toffoli) Gate and Fredkin Gate. In Multi Controlled Toffoli (MCT) from the domain of discourse, containing $n$-variables, $(n-1)$ variables are treated as the control inputs and $1$ variable is the target output which will be inverted iff all control lines are assigned to $1$. If number of control inputs $(n-1)$ is $2$, then MCT is called Toffoli Gate, and if $(n-1)$ is $1$, then MCT is called Controlled NOT (CNOT) gate. Fredkin gate has $n$ control lines and $r$ target lines which will be interchanged iff the conjunction of all $n$ input lines evaluates to $1$. If a Fredkin gate does not have any control input, then it is called a SWAP gate. Quantum physical circuit architecture, invented so far, can process only single qubit and two-qubit gates. Implementation of multi-controlled quantum gates into physical circuits is not feasible. So, decomposition of complex gates into a sequence of elementary quantum gates like NOT (Inverter), CNOT (Controlled-NOT), Controlled-$V$, Controlled-$V^{+}$(Inverse of Controlled-V) etc are required.
\begin{figure}
\begin{center}
\includegraphics[width=\textwidth]{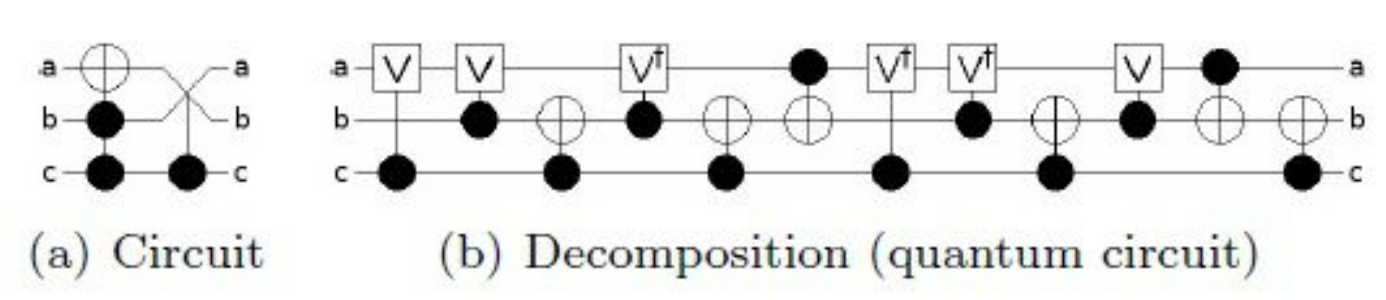}
\end{center}
\caption{Decomposition of Toffoli and Fredkin gates} \label{ToffoliFredkin}
\end{figure}
Figure ~\ref{ToffoliFredkin} shows decomposition of Toffoli and Fredkin gates into elementary CNOT, Controlled-$V$ and Controlled-$V^{+}$gates. During synthesis step involved in physical realization of Quantum circuits, optimization of circuit levels as well as gate count in a quantum boolean network needs to be done \cite{WilleReversible}. Proper algorithmic approach is required for minimizing quantum cost of the circuit. Several performance metrics include:
\begin{itemize}
\item Number of lines and constant inputs: Initialization of quantum registers is complex because of the exponential state-space of an $n$-qubit register.
\item Gate count and Quantum cost: Number of elementary quantum operations needed to realise a gate contributes to quantum cost.
\item Circuit Depth: Number of steps required to execute all available gates in a circuit.
\item Gate Distribution: Coherence time for qubits and operation time for gates are widely affected by technological parameters as the total operation time of gates applied to a qubit must never exceed its qubit decoherence time. Otherwise the qubit value is lost before applying all gates \cite{WilleReversible}.

\item Nearest-Neighbor Cost: Most promising performance metric involved in physical realization of quantum circuit is the NNC (Nearest-Neighbor Cost). In real quantum technologies some restrictions exists between two interacting qubits. Most of the physical implementations follow Linear Nearest-Neighbor (LNN) architecture where two qubits are allowed to interact if and only if they are adjacent to each other.

\end{itemize}
\begin{figure}
\begin{center}
\includegraphics[width=300pt]{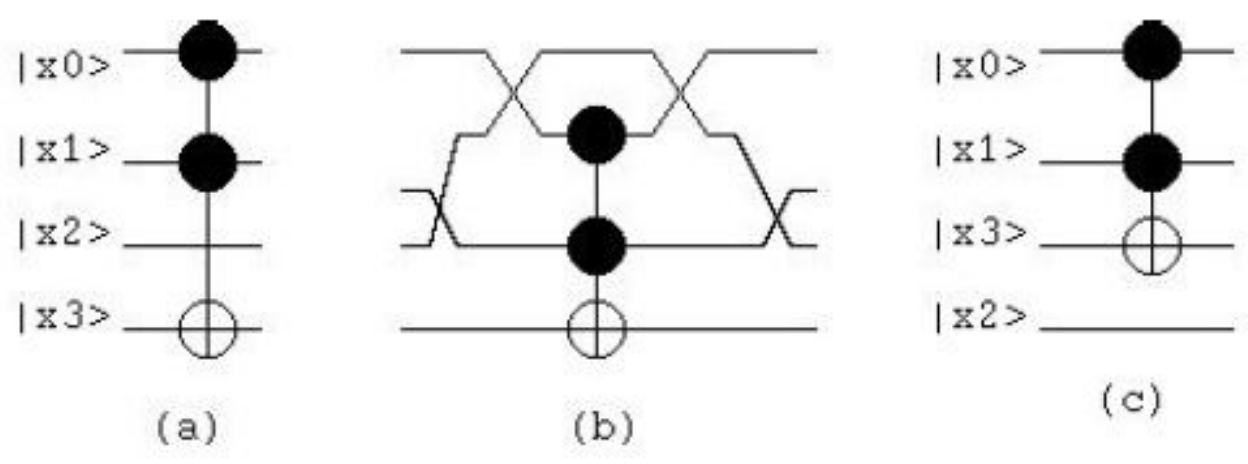}
\end{center}
\caption{Reordering of Toffoli gates for LNN synthesis
(a)Toffoli gate in non-LNN architecture (b) 4 SWAP gates without qubit reordering (c)Reordering circuit with no requirement of SWAP gate} \label{toffoli}
\end{figure}
\subsection{Linear Nearest Neighbor (LNN) Synthesis}
In order to minimize NNC, qubit lines must be reordered so that non-adjacent qubits can be adjacent before interaction. Without loss of generality, it is assumed that a given Quantum Boolean Circuit (QBC) is not in nearest neighbour form. In order to convert a given QBC to a corresponding LNN architecture, SWAP gates must be inserted appropriately. In Figure ~\ref{toffoli}, a Toffoli gate is shown which is not in LNN architecture. If qubit lines are not reordered then number of SWAP gates in LNN representation is not optimal, whereas optimal solutions can be achieved after reordering of qubit registers resulting in less number of SWAP gates \cite{AChakrabartiLNN}. 

LNN synthesis for NOT, CNOT and Toffoli (NCT) and Multi-Controlled Toffoli (MCT) should be handled differently. The number of SWAP gates for a single qubit NOT gate is zero and for that of CNOT gate, it is simply the number of intermediate qubit lines between top and bottom control lines.
\begin{figure}
\begin{center}
\includegraphics[width=300pt]{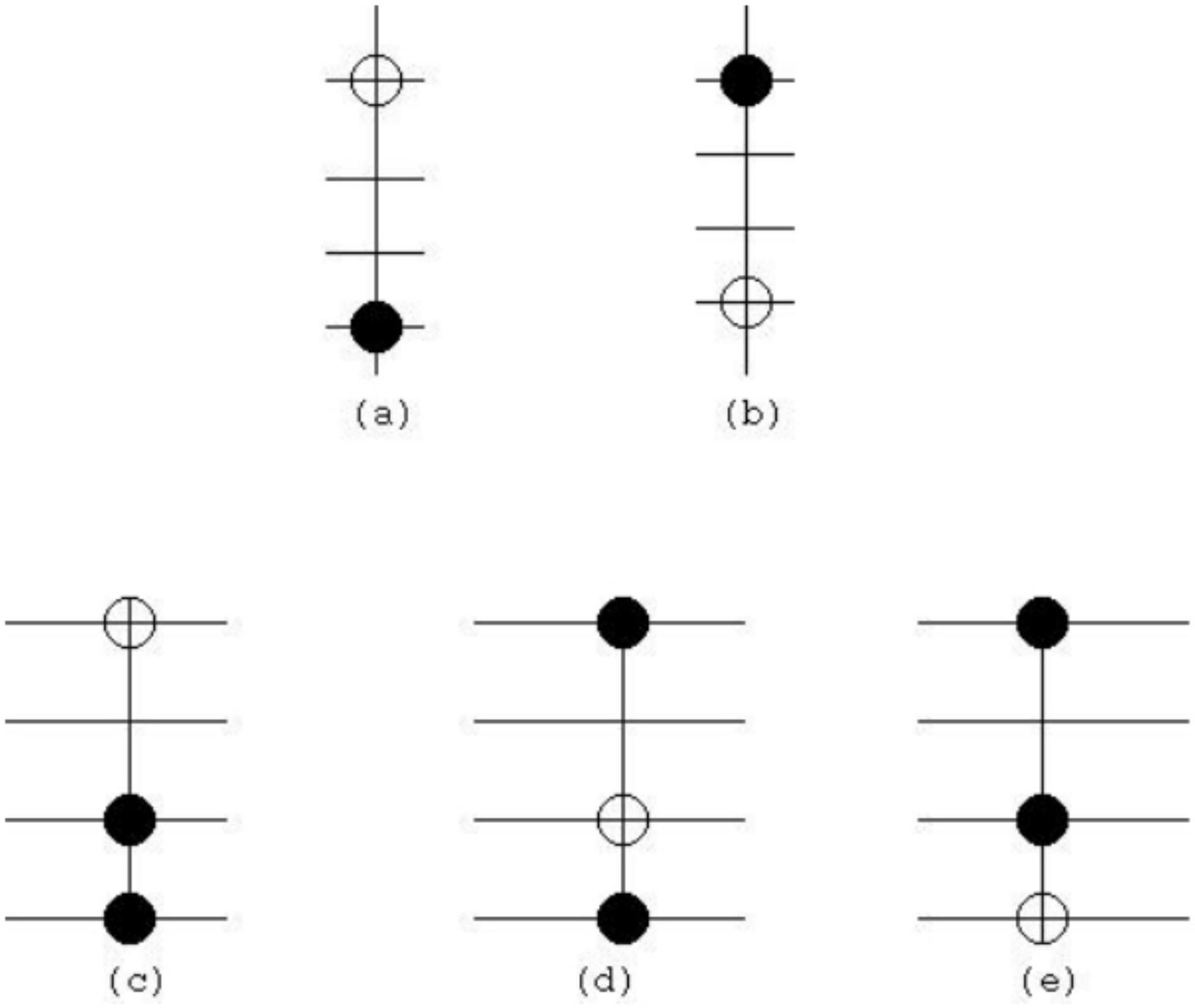}
\end{center}
\caption{Variation of MCT
(a) (b)Types of CNOT (NCT) (c)(d)(e) Types of TOFFOLI
} \label{MCT}
\end{figure}

Decomposition of Multi-Controlled Toffoli gates results in increased number of SWAP gates. For decomposition of a single $C^{k}$ NOT gate, the number of TOFFOLI required is $2(k-1)+1$ and number of auxiliary qubits is $(k-2)$ \cite{AChakrabartiLNN}. 

\begin{figure}
\begin{center}
\includegraphics[width=200pt]{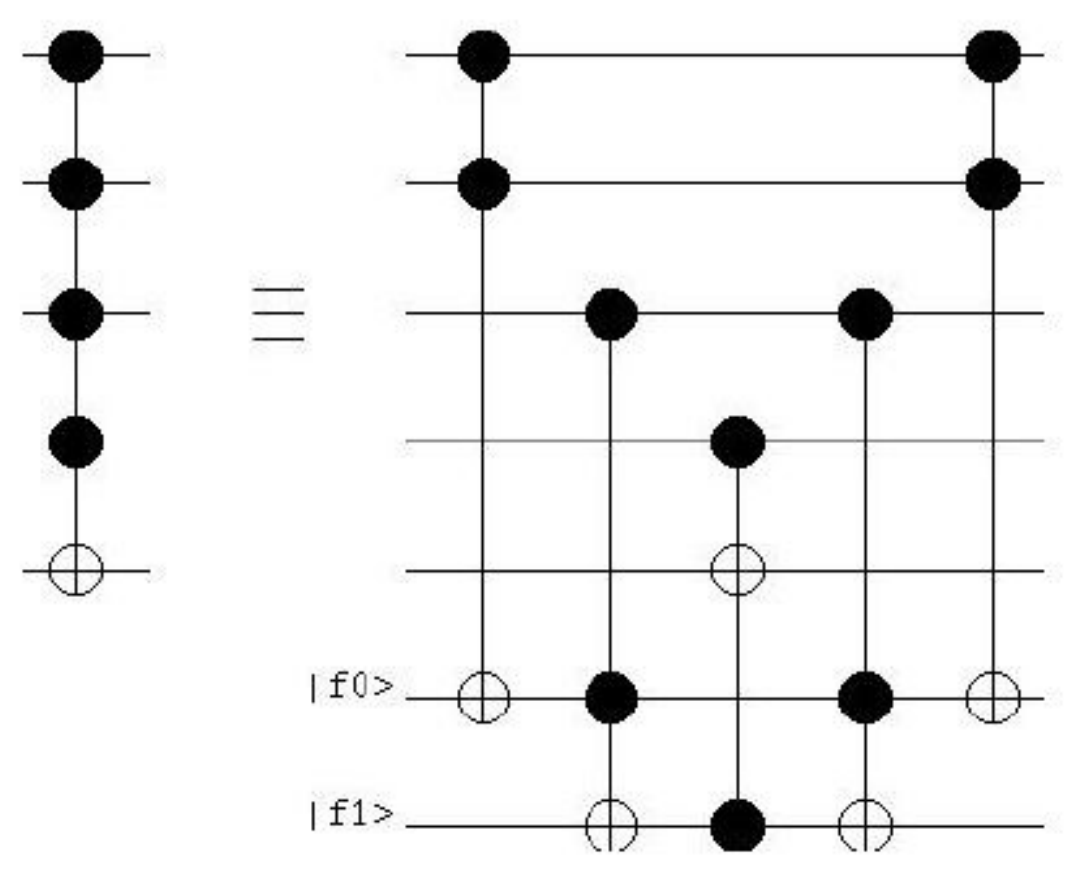}
\end{center}
\caption{Decomposition of MCT gate: replacement of a $C^{4}$NOT gate by equivalent Toffoli (NCTs) with two auxiliary qubits $f_{0}, f_{1}$.} \label{MCTDecomposition}
\end{figure}
The general idea of NNC optimization is to apply adjacent SWAP gates whenever a non-adjacent quantum gate occurs in the standard decomposition. SWAP gates are added in front of each gate with non-adjacent control and target lines to move a control line of a gate towards the target line until they become adjacent. In order to restore the original ordering of circuit lines, SWAP gates should be also added afterwards. A quantum gate 'g' where its control and target are placed at $a^{th}$  and $b^{th}$ lines, additional quantum cost of $x.\mid a-b-1 \mid$ is needed. $\mid a-b-1 \mid$ number of adjacent SWAP operations are required in order to make qubits adjacent, where x is the quantum cost for one SWAP operation \cite{WilleReversible}. \cite{AChakrabartiLNN} In order to minimize the number of SWAP gates, placement of qubits and establishment of a routing channel are two essential stages. In 2-Dimensional Nearest-neighbor, Two-qubit gate, Concurrent (NTC) architecture, any arbitrary circuit can be represented in a placement grid where maximum number of adjacent positions corresponding to a given cell is four. The goal of qubit placement is to place highly interactive qubits nearby in that grid and this can be done with the help of an interaction graph, where the vertices refer to qubits and edge weights refer to the number of two-qubit gates operating on two-qubits. Consecutive gates in a given circuit can be executed in parallel due to sharing of control and target qubits which results in a working set of very few gates of a circuit at one scheduling level. This idea was incorporated by A. Shafaei et. al. through providing a solution where the first phase starts with the formulation of  $m$ instances of the grid-embedding space on $m$ sub circuits of an interaction graph for a circuit with $N$ gates. The next phase is to insert SWAP gates before non-local two qubit gates of each sub circuit to obtain final placement. The final step requires a swapping network to align qubit arrangements of two consecutive sub circuits using 2-D bubble sort algorithm\cite{ShafaeiPlacement2D}. Another efficient qubit placement algorithm was proposed by Chia-Chun Lin et. al. where their algorithm was based on breadth-first traversal. \cite{PAQCS} The corresponding inputs to the logic circuits are the logic circuit and a parameter used as ranking of qubits to be chosen for their placement in neighboring cells of a given qubit. In their work, they had taken degree of a vertex and activity of a vertex which is summation of all edge weights of its neighboring qubits into consideration for determining priority of a qubit over other qubits. After selection of a vertex is over, that vertex is placed in the placement grid \cite{PAQCS}.
Now, say a vertex $q_{0}$ has more than four adjacent nodes $q_{1}, q_{2},q_{3}, q_{4}, q_{5} $ and considering the grid to be vacant four high-priority nodes  $q_{1}, q_{2},q_{3}, q_{4}$ are placed on the grid at four neighboring positions. There exists another node say $q_{6}$ which has adjacency with  $q_{5}$ and rank of  $q_{6}$ is $k_{6}$ and $k_{1}>k_{6}>k_{i,i=24}$. Since, node $q_{0}$ is deleted from priority queue so $q_{5}$ cannot be placed in adjacent positions in first iteration. So during the next iteration $q_{6}$ can never be placed though it has higher priority over adjacent nodes $q_{1}, q_{2},q_{3}, q_{4}$.Thus, number of SWAP gates will increase as direct interaction between two adjacent nodes cannot be kept as that of logical circuit in the physical implementation.
A different scenario can also happen during the physical implementation phase. The rank of a vertex (qubit) is determined by a function $f$ which takes into account both the activity and degree of a node. Say, vertex $q_{0}$ is chosen first as its $f$ is maximum. Now, consider a situation where a vertex $q_{2}$ is chosen with second maximum value where activity value is 100 and degree is 50, making f=150 but there exists another vertex $q_{1}$ whose degree is 2 but activity value is 147 making f=149. Since $q_{2}$ is chosen after $q_{0}$, so $q_{1}$ can never be chosen if $q_{1}$ is adjacent to only $q_{0}$ and there is no vacant position in the neighboring cells of $q_{0}$ in the grid.
In our work, we have proposed a technique to find the long path of highly interactive qubits where selection of vertices (qubits) will be made such that optimization in terms of selection of desired qubits can be made in order to provide reliability in the physical synthesis cost of any quantum circuit.

\section{Proposed approach}
In this section, we are proposing two novel algorithms to optimize the number of SWAP gates based on edge weight optimization and removal of pair operations from the net-list respectively. Given a quantum circuit Q, in its QASM form, we are looking for a optimal insertion of SWAP gates so that all gates g of Q can be executed in adjacent manner.

\begin{algorithm}[H]
\footnotesize
\SetAlgoLined
\KwResult{A $n \times n$ grid, where all the qubits of a quantum circuit are placed by their index.}
\textbf{Input :} QASM file $Q$, for a quantum circuit after FTQLS\;
\textbf{Initialization:} set $q = $number of qubits in the quantum circuit\;
set $G[i][j]=0, \forall i,j=1\dot{...}q$, set $path[i]=0, \forall i=1\dot{...}q$ and set $maxdeg=-1$\;
set $n=\sqrt{q}$, and set $grid[i][j]=-1, \forall i,j=1\dot{...}n$\;
Read $Q$ \While{$!EOF$}{
  for each line $L$ in $Q$\;
  \eIf{number of qubits in $L == 2$}{
   find index $i$ and $j$ of the two qubits $\in L$\;
   set $G[i][j] = G[i][j] + 1$\;
   }{
   Discard $L$\;
  }
 }
 set $i=0$, $j=0$ and $max=0$\;
 \While{$i<q$}{
 \While{$j<q$}{
	\eIf{$G[i][j]!=0$}{
		$path[i] = path[i] + 1$ and $path[j] = path[j] + 1$\;		   
   }{$Continue$\;} 
 }
 }
 set $j=0$\;
 \While{$i<q$}{
	\eIf{$path[i]>j$}{
		set $j=path[i]$ and set $maxdeg=i$\;	
	}{$Continue$\;}
 }
set $path[i]=-1, \forall i=1\dot{...}q$, set $path[0]=maxdeg$, and set $i=1$\;
\While{$i<q$}{
set $j=-1$ and set $selected=$column index $j$ of the element with maximum value among row $G[i-1]$, where $j \notin path$ \;
\eIf{$j!=-1$}{
set $path[i]=selected$ and set $G[path[i-1][path[i]]=G[path[i][path[i-1]]=0$\;
}
{set $path[i]=j$, where $j$ is the row index of maximum value in $G$ and $j \notin path$\; }
set $i=i+1$\;
}
set $path[0]$ at $grid[\mid (n-1)/2 \mid][\mid (n-1)/2 \mid]$ and the generated long path \textbf{spirally} to the $grid[n][n]$ matrix\;
\caption{\textbf{setLongPath() - generates interaction graph for a given quantum circuit and finds a long path to set the path in a grid}}
\par
\end{algorithm}

\begin{algorithm}[H]
\footnotesize
\SetAlgoLined
\KwResult{QASM file Q, with optimally inserted SWAP gates.}
\textbf{Input :} QASM file $Q$, for the quantum circuit after FTQLS and $grid[n][n]$\;
\textbf{Initialization:} set $q = $number of qubits in the quantum circuit\;
Read $Q$ \While{$!EOF$}{
  for each line $L$ in $Q$\;
  \eIf{number of qubits in $L == 2$}{
   find index $i$ and $j$ of the two qubits $\in L$\;
   find $x1$ and $y1$, where $grid[x1][y1]=i$\;
   find $x2$ and $y2$, where $grid[x2][y2]=j$\;
   Insert SWAP gate before $L$ as follow (Considering $x1<=x2$ and $y1<=y2$):
   set $i=x1$\;
   \While{$i<=x2$}{
   	INSERT SWAP($grid[i][y1],grid[i+1][y1]$) before $L$\;
	$i=i+1$\;   	
	}
	set $i=y1$\;
	\While{$i<y2$}{
   	INSERT SWAP($grid[x2][i],grid[x2][i+1]$) before $L$\;
	$i=i+1$\;   	
	}
   }{
   Discard $L$\;
  }
 }
 Read $Q$ \While{$!EOF$}{
 for each line $L1$ in $Q$\;
 check the next line $L2$ in $Q$\;
 \eIf{$L1$ is identical to $L2$}{
Goto the previous line $L0$ of $L1$\;
Remove both the lines $L1$ and $L2$ from $Q$\;
set $L1=L0$\;
 }{$Continue$\;}
 }
\caption{optimizeRoute() - optimize the number of SWAP gates during the SWAP operations}
\end{algorithm}

The First algorithm  \textbf{(setLongPath())} is generating an interaction graph from a given quantum circuit. The generated graph will denote the number of qubits of the given circuits as its set of vertices, the operations between different qubits as its set of edges and the number of binary operations between any two qubits as the weight of every individual edge.Then the algorithm takes the generated graph to find a long path starting with the node, having maximum degree. Then it sets the generated long path into a $n \times n$ grid. The starting node of the path, having maximum degree will be assigned to $(\mid (n-1)/2 \mid, \mid (n-1)/2 \mid)^{th}$ cell of the grid. If all the vertices are not present in the long path, then the algorithm takes remaining vertices to generate next long path in order to append it with the previous output. The placement of qubits in their respective cells of the grid will generate a long path in spiral fashion. Hence, the initial placement of qubits in a 2D architecture is done by this algorithm to optimize the number of SWAP gate insertions. The algorithm considers greedy approach to include new nodes into the long path. It checks the interactions between every pair of qubits in order to find the maximum interacting qubits among them. Our initial assumption is that the number of qubits is ‘q’. Hence, the complexity of the first algorithm is $O(q^2)$, where $q$ is the number of qubits in the quantum circuit.

The second algorithm \textbf{(optimizeRoute())} is designed to optimize the number of SWAP gates during routing step to achieve any quantum operation between two non-neighboring qubits. It takes Quantum Assembly (QSAM) file with quantum gate list and the generated grid as input and inserts SWAP gates at the required position. Then it finds all the pair of consecutive gate operations, where the operation and operand(s) are same and removes that pair of operations from the QASM file. Hence the gate cost is optimized. The algorithm reads each instruction from the generated QASM file and apply
SWAP gates as needed. Therefore, the complexity of \textbf{optimizeRoute()} is $O(I)$,
where $I$ is the number of instructions in the generated QASM file by our first
algorithm.

\section{Proof of correctness}

Consider an interaction graph $G$ consisting of sets of vertices (qubits), sets of edges (quantum operations) and respective edge weights (number of SWAP gates) is given below. Our placement algorithm setLongPath() chooses vertex ,$q_{5}$ as the initial vertex as its degree of interaction is maximum.
 
\begin{figure}
\begin{center}
\includegraphics[width=300pt]{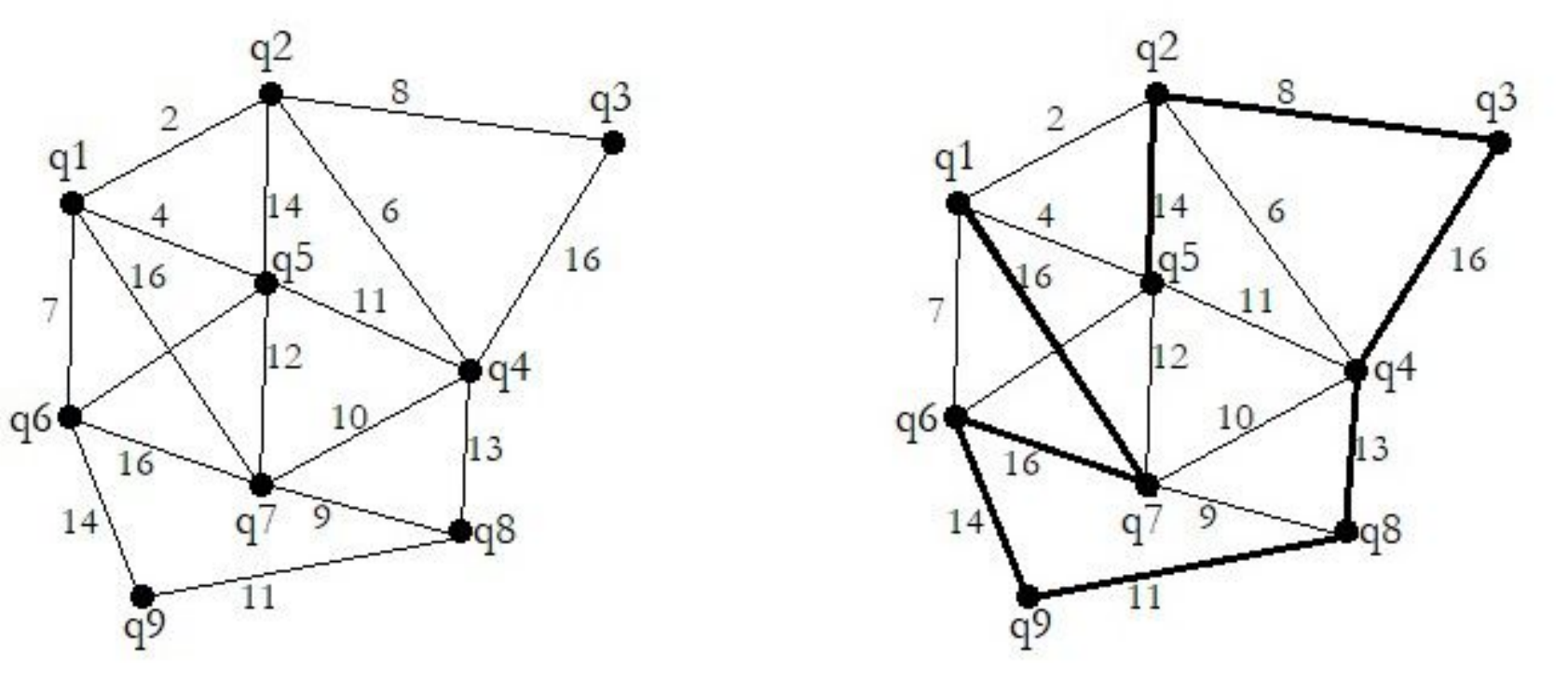}
\end{center}
\caption{ Interaction Graph of a quantum circuit and the long path on it
} \label{InteractionGraph}
\end{figure}

Then a greedy approach is applied to select the next vertex among the neighboring vertices of the chosen vertex with the maximum edge weight as it represents maximum interaction in terms of requirement of SWAP gates. 

\begin{figure}
\begin{center}
\includegraphics[width=250pt]{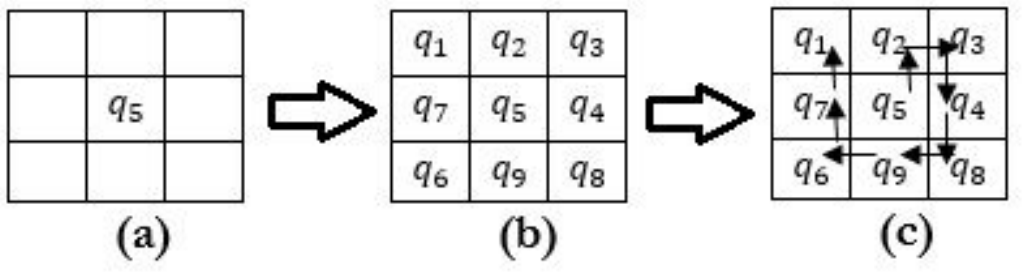}
\end{center}
\caption{Steps involved in placements of qubits in an $n \times n$ placement grid
(a) Placement of initial vertex, (b) Placement of all vertices, and (c) Selection of Long path using setLongPath() algorithm. Hence the initial placement of qubits is done by our proposed first algorithm.
} \label{setLongPath}
\end{figure}

The output of the algorithm will be an optimal path ensuring the reachability of all vertices (qubits) giving priority to highly interacting qubits. This will give us an optimal path ($q_{5}-q_{2}-q_{3}-q_{4}-q_{8}-q_{9}-q_{6}-q_{7}-q_{1}$)which can be easily placed on a $n \times n$ grid where qubit placement will be stored from ($\mid (n-1)/2 \mid , \mid (n-1)/2 \mid$)th position of the grid and placement of remaining qubits will be done in spiral fashion giving us a linear path as shown in Figure ~\ref{setLongPath}.

\begin{figure}
\begin{center}
\includegraphics[width=280pt]{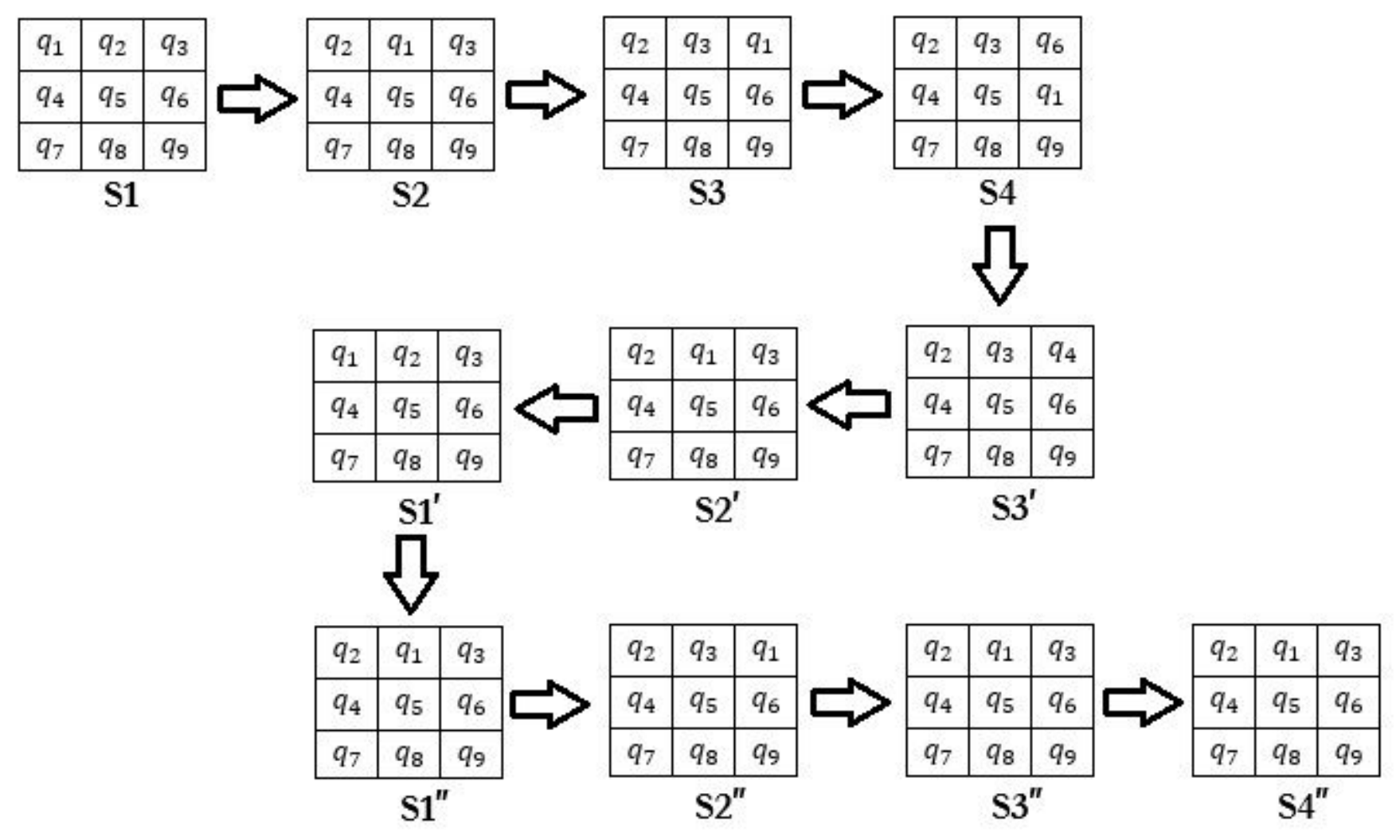}
\end{center}
\caption{Placement of qubits during routing in $n \times n$ grid
} \label{placement}
\end{figure}
\begin{figure}
\begin{center}
\includegraphics[width=230pt]{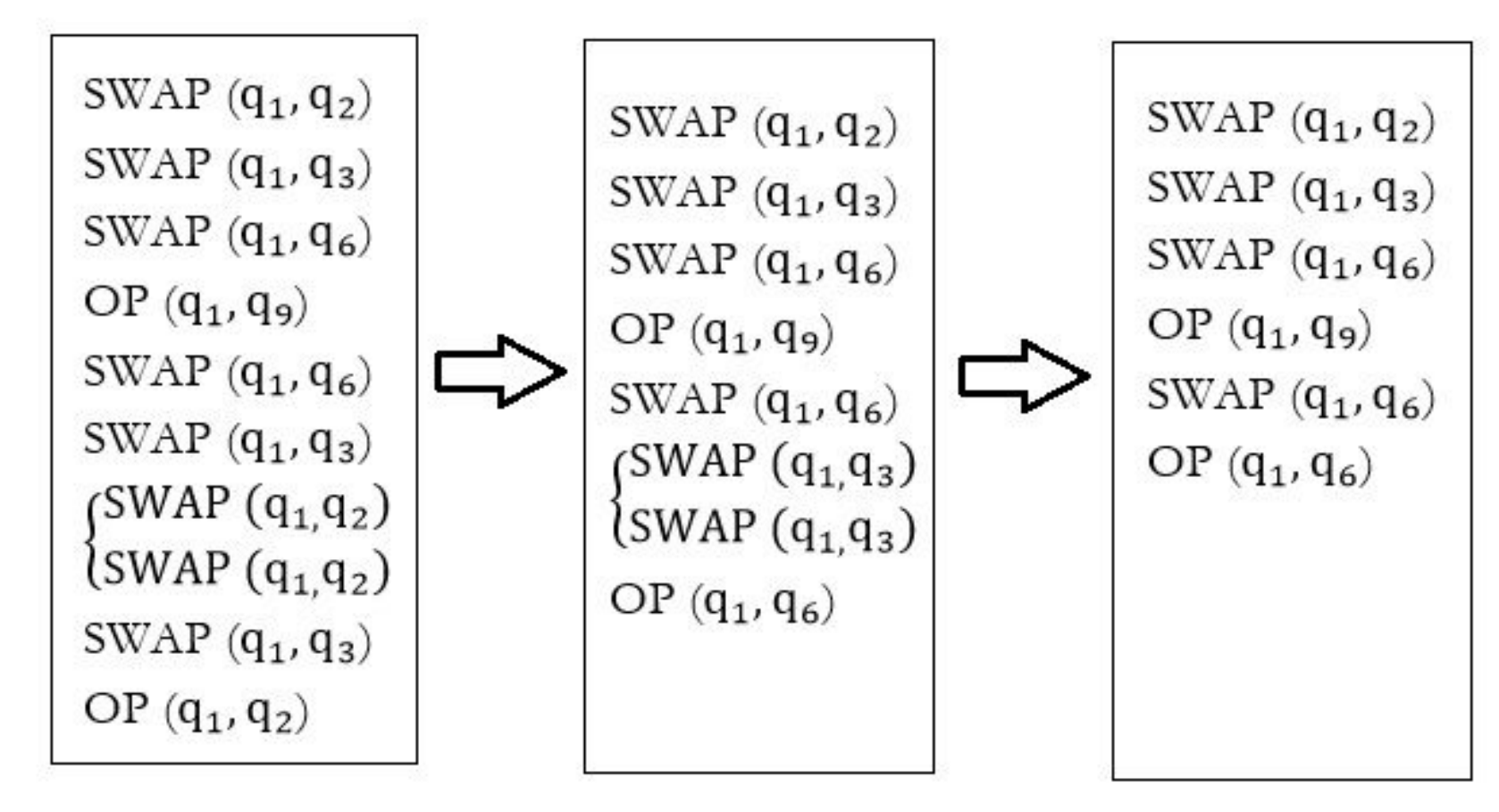}
\end{center}
\caption{Reduction in number of steps involved in OptimizeRoute() algorithm to reduce the number of SWAP gates to be inserted.
} \label{ReduceSWAP}
\end{figure}

The initial grid configuration is represented in step $S1$. Steps $S2$, $S3$ and $S4$ are required to make non-adjacent qubits $q_{1}$, and $q_{9}$ adjacent to each other.Steps $S3^{\prime}$, $S2^{\prime}$ and $S1^{\prime}$ are required to retain the initial configuration once operation between qubits $q_{1}$ and $q_{6}$ is performed. Now, if another quantum operation is defined between $q_{1}$ and $q_{6}$, then a communication channel is to be made up through steps $S1^{\prime\prime}$ to $S3^{\prime\prime}$. The final configuration is shown in steps $S4^{\prime\prime}$. But, this placement leads to non-optimal routing as it requires four additional SWAP gates which can be avoided through our algorithm optimizeRoute() as it removes unnecessary intermediate SWAP gates as show in Figure ~\ref{ReduceSWAP}. Thus, an optimized routing is obtained in physical synthesis of the given quantum circuit.

\section{Conclusion}
In this work, we have discussed the issues regarding implementation of methodological framework of physical quantum circuit synthesis. The necessity of incorporating SWAP gates is also mentioned so as to make a non-LNN architecture behave as an LNN one. Our proposed algorithms setLongPath() and optimizeRoute() have led to significant reduction in the number of SWAP gates required and can be applied over any arbitrary quantum circuit. We will attempt to extend our research for achieving cost optimization in 3D NTC architecture.

\end{document}